\documentclass[twocolumn]{aastex631}

\usepackage{float}
\usepackage{amsmath}
\usepackage{soul}

\newcommand{\msun}{\mathrm{M_{\odot}}} 

\newcommand{\mhe}{M_{\mathrm{He}}} 
\newcommand{\mwd}{M_{\mathrm{WD}}} 
\newcommand{\mch}{M_{\mathrm{Ch}}} 

\newcommand{\lsun}{\mathrm{L_{\odot}}} 

\newcommand{\rsun}{\mathrm{R_{\odot}}} 
\newcommand{\Rhe}{R_{\mathrm{He}}} 

\newcommand{\iniMhe}{M^{i}_{\mathrm{He}}}

\newcommand{\iniPc}{P^{i}_{\mathrm{c}}}
\newcommand{\inirhoc}{\rho^{i}_{\mathrm{c}}}

\newcommand{\mesa}{{\tt\string MESA}}
\newcommand{\athena}{{\tt\string Athena++}}

\newcommand{\kB}{k_{\mathrm{B}}}
\newcommand{\NA}{N_{\mathrm{A}}}

\newcommand{\gcc}{\mathrm{g} \, \mathrm{cm}^{-3}}
\newcommand{\kms}{\mathrm{km}\,\mathrm{s}^{-1}}

\newcommand{\Porb}{P_{\mathrm{orb}}}

\newcommand{\Pc}{\ensuremath{P_{\rm c}}}
\newcommand{\rhoc}{\ensuremath{\rho_{\rm c}}}

\newcommand{\Sc}{s_{c}}

\newcommand{\vel}{\textbf{v}}

\newcommand{\aext}{\textbf{a}_{\mathrm{ext}}}
\newcommand{\gacc}{\textbf{g}_{\mathrm{acc}}}
\newcommand{\acent}{\textbf{a}_{\mathrm{cent}}}

\newcommand{\Be}{\mathrm{Be}}

\newcommand{\Mej}{M_{\mathrm{ej}}}
\newcommand{\Eej}{E_{\mathrm{ej}}}
\newcommand{\Eke}{E_{\mathrm{KE}}}

\begin{document}

\title{Shocking and Mass Loss of Compact Donor Stars in Type Ia Supernovae}

\author[0000-0001-9195-7390]{Tin Long Sunny Wong}
\affiliation{Department of Physics, University of California, Santa Barbara, CA 93106, USA}

\author{Christopher J. White}
\affiliation{Center for Computational Astrophysics, Flatiron Institute, Simons Foundation, New York, NY 10010, USA}
\affiliation{Department of Astrophysical Sciences, Princeton University, Princeton, NJ, USA}

\author[0000-0001-8038-6836]{Lars Bildsten}
\affiliation{Department of Physics, University of California, Santa Barbara, CA 93106, USA}
\affiliation{Kavli Institute for Theoretical Physics, University of California, Santa Barbara, CA 93106, USA}

\correspondingauthor{Tin Long Sunny Wong}
\email{tinlongsunny@ucsb.edu}

\begin{abstract}
Type Ia supernovae arise from thermonuclear explosions of white dwarfs accreting from a binary companion. 
Following the explosion, the surviving donor star leaves at roughly its orbital velocity. 
The discovery of the runaway helium subdwarf star US 708, and seven hypervelocity stars from Gaia data, all with spatial velocities $\gtrsim 900$\,km/s, strongly support a scenario in which the donor is a low-mass helium star, or a white dwarf. 
Motivated by these discoveries, we perform three-dimensional hydrodynamical simulations with the {\texttt{Athena++}} code modeling the hydrodynamical interaction between a helium star or helium white dwarf, and the supernova ejecta. 
We find that $\approx 0.01-0.02\,M_{\odot}$ of donor material is stripped, and explain the location of the stripped material within the expanding supernova ejecta. 
We continue the post-explosion evolution of the shocked donor stars with the {\texttt{MESA}} code. As a result of entropy deposition, they remain luminous and expanded for $\approx 10^{5}-10^{6}$~yrs. We show that the post-explosion properties of our helium white dwarf donor agree reasonably with one of the best-studied hypervelocity stars, D6-2.

\end{abstract}


\section{Introduction}

Although type Ia supernovae (SNe Ia) are important cosmological distance indicators \citep[e.g.,][]{Riess1998,Perlmutter1999}, the identity of their progenitor systems remains debated today. There is general agreement that SNe Ia result from the thermonuclear explosion of a carbon-oxygen white dwarf \citep[C/O WD; ][]{Hoyle1960} induced by accretion from a companion donor star, but the stellar type of the donor remains elusive \citep[see, e.g.,][for a review]{Hillebrandt2000,Maoz2014,Liu2023_review}. 

In the Chandrasekhar-mass scenario, the accretor grows up to Chandrasekhar mass and explodes, through accretion from a non-degenerate companion \citep[][]{Whelan1973,Nomoto1982}. 
However, the non-detection of a progenitor system in pre-explosion images of nearby SNe \citep[e.g., ][]{Li2011,Kelly2014}, the lack of uncontested surviving companions in galactic and Large Magellanic Cloud SN Ia remnants \citep[e.g.,][]{RuizLapuente2004,Kerzendorf2009,Schaefer2012,Kerzendorf2014,Kerzendorf2018,RuizLapuente2018,RuizLapuente2019,Shields2023}, etc., have dimmed the prospects of this scenario.

In the double-detonation scenario, the accretor WD accumulates $0.01 - 0.1\,\msun$ of helium (He) from a He-rich companion. 
The accumulated He shell detonates and sends a shock wave into the the C/O core leading to its subsequent detonation \citep[e.g.,][]{Nomoto1982,Woosley1986,Livne1990,Livne1991,GarciaSenz1999,Fink2007,Fink2010,Kromer2010,Woosley2011,Pakmor2012,Sim2012,Moll2013,Shen2014,Polin2019,Townsley2019,Gronow2020,Gronow2021,Leung2020,Boos2021}. 
This scenario requires the donor to be a Roche-lobe-filling He star \citep[][]{Iben1991,Brooks2015,Bauer2017,Neunteufel2019}, He WD \citep[e.g.,][]{Bildsten2007,Burmester2023,Wong2023}, or C/O white dwarf with a surface He layer \citep[e.g.,][]{Guillochon2010,Dan2011,Pakmor2012,Pakmor2013,Dan2015}. 
The orbital period at explosion is $\lesssim 15$~minutes, and the donor orbital velocity is $v_{\rm orb} \gtrsim 600\,\kms$ ($\gtrsim $~1,000~$\kms$ for a white dwarf donor). 
When the SN explosion unbinds the binary, the surviving donor continues at its velocity $v \approx v_{\rm{orb}} \gtrsim 600\,\kms$. 

Discovery of the runaway He-burning subdwarf star US~708 \citep[$v\approx900-$1,000$\,\kms$;][]{Geier2015,Neunteufel2020,Heber2023}
and seven hypervelocity WDs \citep[$v \approx $~1,000$-$2,000$~\kms$;][]{Shen2018_D6,ElBadry2023} provides smoking gun evidence that the double-detonation scenario can account for some SNe Ia. 
The hypervelocity WDs appear to have larger radii ($\approx 0.02-0.2\,\rsun$) compared to typical white dwarfs, likely due to pre-explosion tidal-heating, or shock-heating by the SN ejecta \citep[][]{Shen2018_D6,ElBadry2023,Bauer2019}. Some have metal-polluted atmospheres, likely due to contamination by the SN ejecta \citep[][]{Shen2018_D6,ElBadry2023}. 
However, note that no high-proper-motion object brighter than $L > 0.0176 L_{\odot}$ was found in the remnant of SN 1006 by \cite{Shields2022}.

Previous numerical studies of SN ejecta-companion interaction have largely focused on non-degenerate companions relevant to the Chandrasekhar-mass scenario, including a main sequence star, subgiant, red giant, and He star \citep[e.g.,][]{Marietta2000,Pakmor2008,Pan2010,Pan2012a,Liu2013a,Liu2013b,Liu2013c,Boehner2017,Zeng2020,Rau2022,McCutcheon2022}. In the context of the double-detonation scenario, only \cite{Bauer2019} \& \cite{Liu2021a} studied the response of a He star to interaction with SN ejecta and its long-term evolution, and recently \cite{Bhat2024} studied the long-term evolution of CO WD donors. On the other hand, \cite{Papish2015}, \cite{Tanikawa2018,Tanikawa2019}, \cite{Pakmor2022}, \cite{Burmester2023}, \cite{Boos2024} \& \cite{Shen2024} found that in some cases, the donor also detonates and leaves behind no surviving remnant.

We continue the work of \cite{Bauer2019} and model the interaction between SN ejecta and He WD and He star donors, using the {\athena} hydrodynamic code \citep[][]{Stone2020}. 
We improve upon their work by using a new passive scalar capability to differentiate between donor and SN ejecta materials, and adopting a more realistic SN ejecta profile based on the sub-$\mch$ explosion simulations by \cite{Shen2018_subMch}.  In addition to the two He star models first introduced in \cite{Bauer2019}, we also present a low-mass, high-entropy (semi-degenerate) He WD model from \cite{Wong2023}. 
They modeled the stable mass transfer from a high-entropy He WD donor onto a C/O WD accretor, up to the start of a dynamical He flash on the accretor that yields a He detonation. Assuming the donor leaves with a spatial velocity $\approx v_{\rm orb}$, they found good agreement with the spatial velocity of the hypervelocity WD D6-2 \citep[][]{Bauer2021}.

Our paper is organized as follows. 
In Section \ref{sec:numerics}, we describe our {\athena} setup. In Section \ref{sec:ejecta}, we describe our SN ejecta profile based on simulations by \cite{Shen2018_subMch}. We detail the outcomes for a He WD donor with a total ejecta kinetic energy $\Eke = 1.2 \times 10^{51}$~erg and ejecta mass $\Mej = 1.0\,\msun$ in Section \ref{sec:fiducial}, and discuss the observational implications for donor mass loss mixed with SN ejecta. In Section \ref{sec:other_Eke} we compare the results of varying $\Eke$ between $0.5$ and $1.5\times 10^{51}$~erg for the same He WD donor. We then describe the results with He star donors in Section \ref{sec:HeStar}. In Section \ref{sec:postexp}, we map the post-explosion donor from {\athena} to the 1D stellar evolution code {\mesa} \citep[][]{MESAI,MESAII,MESAIII,MESAIV,MESAV,MESAVI} and study its subsequent evolution. We find a reasonable match to the properties of the hypervelocity WD D6-2 given an inferred age of $\approx 10^{5}$~yrs. We conclude in Section \ref{sec:conclusions}. 
Our {\athena} input and output files, simulation movies, and our {\mesa} inlists and outputs are uploaded to Zenodo (\url{https://doi.org/10.5281/zenodo.12850558}).

\section{Numerics}
\label{sec:numerics}

We configure {\athena} to solve $\Gamma=5/3$ ideal Eulerian hydrodynamic equations in Cartesian coordinates, using the second-order van Leer time integrator \citep[][]{vanLeer1979}, piecewise parabolic method for spatial reconstruction \cite[][]{Colella1984}, and HLLC Riemann solver, as was done in \cite{Bauer2019}. In contrast with that work's use of Fourier methods to solve Poisson's equation, here we utilize the new multigrid capabilities of {\athena} for self-gravity \citep{Tomida2023}. The gravitational potential is found via the full multigrid method, using sufficient iterations to reach convergence. We use isolated boundary conditions -- the potential on the grid boundary agrees with the distribution of mass, assuming vacuum outside the domain, up to hexadecapole order about the center of mass. In the fluid sector, we impose diode boundary conditions (zero gradient in density, pressure, and velocity, with velocity outside the domain floored to never point inward); these are overridden with the modeled ejecta where it enters the grid. 

We consider three donor models from the stellar evolution code $\mesa$ \citep[][]{MESAI,MESAII,MESAIII,MESAIV,MESAV,MESAVI} -- a He WD from \cite{Wong2023}, and two He star models from \cite{Bauer2019}. The He WD has an initial mass of $0.21\,\msun$ and a high central specific entropy of $s/(\NA \kB) = 4.0$, where $\NA$ is Avogadro's number and $\kB$ is the Boltzmann constant. It is evolved with an initially $1\,\msun$ CO WD in a binary until a dynamical He flash occurs on the CO WD accretor. At this moment, the He WD mass is $\iniMhe=0.126~\msun$. This model is chosen since it is nearly nondegenerate and its pre-explosion orbital velocity agrees with the velocity of the runaway object D6-2 \citep[][]{Shen2018_D6,Bauer2019,ElBadry2023}. We refer to \cite{Bauer2019} for details of the He star models. However, we stress that all donor models are only semi-degenerate, which makes our assumption of $\Gamma = 5/3$ somewhat reasonable. We further explore the impact of the equation of state in Appendices \ref{appendix_B} \& \ref{appendix_C}.

Following \cite{Bauer2019}, for each of the three donors we adopt their initial central density $\inirhoc$ and pressure $\iniPc$ as the respective simulation units. We also choose $t_{0} \equiv \sqrt{ 5 / ( 8 \pi G \inirhoc ) } $ as the time unit. The velocity and length units follow as $\sqrt{\iniPc/\inirhoc}$ and $x_{0} \equiv \sqrt{\iniPc/\inirhoc} \times t_{0}$. These are provided in Table \ref{tab:models}, where we also provide additional details of each donor model, including the mass $\iniMhe$, radius $R_{\rm He}^{i}$ and sound-crossing time $t_{\rm sound}^{i} \equiv \int ( dr / c_{\rm s} ) $ at the start of the {\athena} simulation, the binary separation $a$, orbital period $\Porb$ and donor orbital velocity $v_{\rm orb}$. 

For all simulations, we set the accretor mass $\mwd$ to be $1\,\msun$ and the binary separation such that the donor is Roche-lobe-filling \citep[][]{Eggleton1983}. All these donor models come from {\mesa} binary simulations where the accretor (albeit with a different mass) is evolved through stable mass transfer up to the start of a dynamical He flash, and so a barely Roche-lobe-filling donor is a reasonable assumption. However, we note that if the double detonation occurs due to unstable mass transfer, the binary separation may be even smaller. For an ejecta mass of $1.0~\msun$ and total kinetic energy $\Eke=1.2\times10^{51}$~erg which defines a peak ram pressure velocity $v_{0}$ (see Section \ref{sec:ejecta}), we define the interaction time $t_{\rm interaction} \equiv a / v_{0}$. We also provide the dimensions of the simulation box in units of $x_{0}$. 
The whole simulation lasts $\approx 150-200\,t_{0}$, until the donor becomes approximately spherical again. 
We set a density and pressure floor of $3 \times 10^{-7}$ and $3 \times 10^{-10}$ in units of $\rhoc$ and $\Pc$. 

\begin{splitdeluxetable*}{cccccccccccBcccc}
\tablenum{1}
\label{tab:models}
\tablecaption{Donor models and description of \athena\, setup}
\tablehead{
\colhead{Donor model} & 
\colhead{$\iniMhe$} & 
\colhead{$R_{\rm He}^{i}$} & 
\colhead{$\mwd$} &
\colhead{$a$} & 
\colhead{$\Porb$} & 
\colhead{$v_{\rm orb}$} &
\colhead{$\rhoc^{i}$} &
\colhead{$\Pc^{i}$} &
\colhead{$x_{0}$} &
\colhead{$t_{0}$} &
\colhead{$t^{i}_{\rm sound}$} &
\colhead{$t_{\rm interaction}$} &
\colhead{Box size ($x$, $y$ \& $z$)} 
\\
\colhead{} &
\colhead{[$\msun$]} &
\colhead{[$\rsun$]} &
\colhead{[$\msun$]} &
\colhead{[$\rsun$]} &
\colhead{[min]} &
\colhead{[$\kms$]} &
\colhead{[$\gcc$]} &
\colhead{[dyne cm$^{-2}$]} &
\colhead{[$\rsun$]} &
\colhead{[s]} &
\colhead{[s]} &
\colhead{[s]} &
\colhead{[$x_{0}$]}
}
\tablewidth{100pt}
\startdata
HeWD & 0.126 & 
0.0410 & 1.0 &
0.186 & 
12.6 & 
954 &
$2.02\times10^{4}$ &
$6.47\times10^{19}$ &
0.00988 &
12.15 &
81 &
14 &
$(-16,64) \times (-50,50) \times(-50,50)$
\\
HeStar1 & 0.348 & 
0.0805 & 1.0 &
0.275 & 20.8 &
717 & $3.74\times10^{4}$ & $2.83\times10^{20}$ & 
0.0112 & 8.93 & 
161 &
21 &
$(-20,60) \times (-40,40) \times(-40,40)$
\\
HeStar2 & 0.236 & 
0.0414 & 1.0 &
0.157 & 
9.34 & 991 & 
$6.71\times10^{4}$ & $4.55\times10^{20}$ &
0.00788 & 6.66 & 
61 &
12 &
$(-15,65) \times (-40,40) \times(-40,40)$
\\
\enddata
\tablecomments{Here $\iniMhe$ and $R^{i}_{\rm He}$ are the initial donor mass and radius, $\mwd$ is the accretor (SN ejecta) mass, $a$ and $\Porb$ are the binary separation and orbital period, $v_{\rm orb}$ is the orbital velocity of the donor, $\rho^{i}_{\rm c}$ and $P^{i}_{\rm c}$ are the initial donor central density and pressure, $x_{0}$ and $t_{0}$ are the length and time units of the simulation, $t^{i}_{\rm sound}$ is the initial donor sound-crossing time, and $t_{\rm interaction} \equiv a/v_{0}$ where $v_{0}$ is given by equation 5.} 
\end{splitdeluxetable*}

\subsection{Passive scalars for ejecta and donor}
\label{sec:scalar}

One novelty of our work compared to \cite{Bauer2019} is that we make use of the new passive scalar capability of {\athena} \citep[][]{Stone2020}. The passive scalar is transported with the fluid without modifying fluid properties. We add two passive scalars, one for the donor and one for the SN ejecta. This allows the accurate tracking of the amount of mass lost from the donor and its velocity distribution. In Section \ref{sec:scalar_results}, we rerun our fiducial simulation but with 4 passive scalars to the donor, so that we can track the original location of mass loss in the donor. This clarifies the interaction between the SN ejecta and the donor. 

\subsection{Donor relaxation to Roche potential}
\label{sec:roche}

Prior to the injection of the SN ejecta, we interpolate the density and pressure profiles of the donor from $\mesa$ onto the {\athena}~grid. We then relax the donor in the corotating frame, accounting for its distortion due to the combined effects of the gravity from the accretor, $\gacc = - \nabla \Phi_{\mathrm{acc}}$, and the centrifugal force, $\acent = - \nabla \Phi_{\mathrm{cent}}$, but without the Coriolis force because it does not contribute to the distortion. The donor potential $\Phi_{\rm d}$ is accounted for in the {\athena} self-gravity module. The other potentials are given by 
\begin{equation}
    \Phi_{\mathrm{acc}} = - \frac{ G \mwd }{ r_{a} } , 
\end{equation}
where $r_{a} = \sqrt{ (x - a)^{2} + y^{2} + z^{2} }$ is the distance from the accretor, and 
\begin{equation}
    \Phi_{\mathrm{cent}} = - \frac{1}{2} \Omega^{2} \sqrt{ ( x - x_{\mathrm{COM}} )^{2} + y^{2} } ,
\end{equation}
where $\Omega = \sqrt{ G ( \mwd + \mhe ) / a^{3} }$ and $x_{\mathrm{COM}} = a [ \mwd / (\mwd + \mhe) ] $. 
We add source terms to the momentum and energy equations as $\rho_{\mathrm{d}} \aext $ and $ \rho_{\mathrm{d}} \vel \cdot \aext $ respectively, where $\aext = \gacc + \acent$ and $\rho_{\mathrm{d}}$ is the density of the donor material (enabled by the use of passive scalars) so that acceleration of the background ``fluff'' does not limit the simulation timestep. 

We apply velocity damping while relaxing the donor to  its Roche configuration. We find this necessary so that (1) large velocities do not severely limit the timestep, (2) the donor does not overshoot its equilibrium structure during relaxation under the Roche potential, and (3) the donor center-of-mass does not change due to imperfect numerical cancellation of the acceleration terms. During relaxation, for each time $t$ with timestep $\Delta t$, the momentum is reduced by a factor $f_{\mathrm{damp}}$ and the kinetic energy $f^{2}_{\mathrm{damp}}$, where 
\begin{align*}
    f_{\mathrm{damp}} 
    &= 
    \exp{ \left( - \frac{ dt_{\mathrm{eff}} }{ \tau_{\mathrm{damp}} } \right) } , 
    \\
    dt_{\mathrm{eff}}
    &= 
    \Delta t \exp{ \left( - \frac{ t }{  t_{\mathrm{damp}} - t  }  \right) }, 
\end{align*}
and $t_{\mathrm{damp}}$ and $\tau_{\mathrm{damp}}$ are taken to be 20 and 1 in units of $t_{0}$. 

\begin{figure*}[]
\centering
\fig{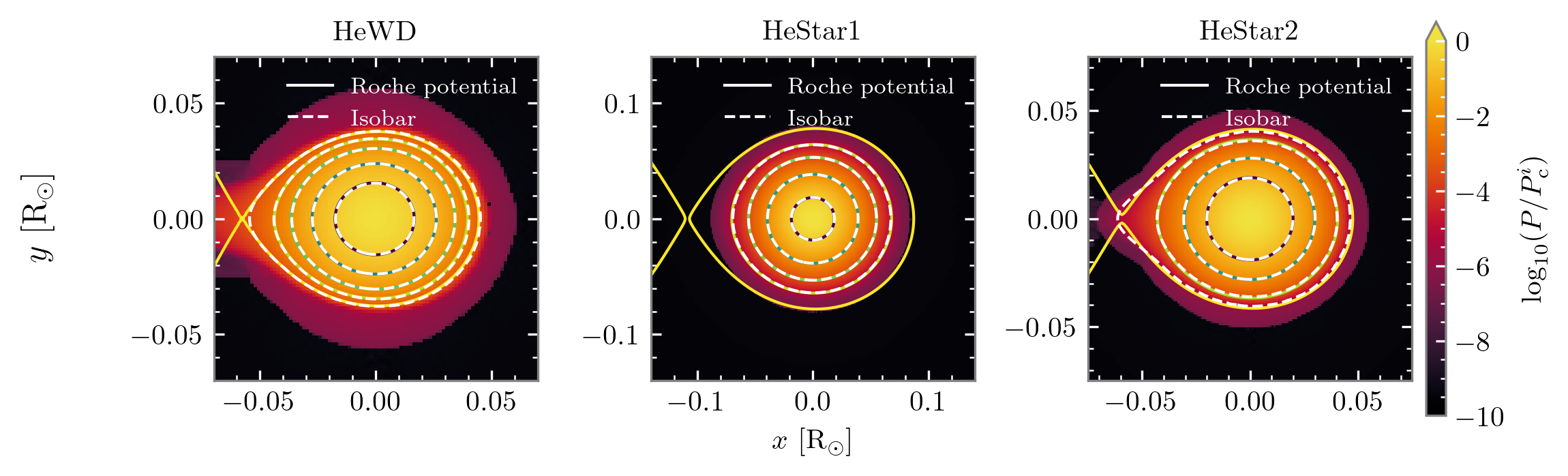}{ 0.95 \textwidth }{}
\caption{ 
Pressure at the mid-plane of the simulation box, right after relaxation is completed at $20\,t_{0}$. White dashed lines show the isobars, whereas colored solid lines show contours of the Roche potential. From left to right, the donors are HeWD, HeStar1 and HeStar2. 
\label{fig:RL}}
\end{figure*}

We confirm that after the relaxation (at 20 $t_{0}$), the isobars of the donor agree well with the equipotentials of the Roche potential $\Phi_{\mathrm{acc}} + \Phi_{\mathrm{d}} + \Phi_{\mathrm{cent}} $
, where $\Phi_{\mathrm{d}}$ is the gravitational potential of the donor calculated by \athena\, since a point-mass approximation is unrealistic. This is shown in Figure \ref{fig:RL}. 

For the HeStar1 model, we find no difference whether we relax the donor with a Roche potential. However, for the HeWD model, the post-explosion central density drops to 10\% of its initial value if we adopt a spherical donor (as opposed to 20\%). We speculate that this difference stems from the detailed interaction between the SN ejecta and the upstream side of the donor especially near the inner L1 Lagrange point, and its effect increases with more extreme mass ratios due to stronger distortion of the donor. 

\subsection{Transitioning the gravitational model as supernova ejecta is introduced}

After relaxation to the Roche configuration (after $20\,t_{0}$), we inject SN ejecta into the grid. At the same time, assuming that the binary is unbound immediately, we switch to the initial donor rest frame. We shut off source terms corresponding to the accretor gravity and the centrifugal force in the corotating frame. 
The assumption that the binary is unbound immediately is valid because the ejecta mass enclosed within the binary rapidly drops within 2~$t_{0}$, and this time is $\ll t^{i}_{\rm sound}$. 

While the unshocked parts of the donor would re-adjust given the absence of the external source terms, we find that the re-adjustment timescale is longer than the shock travel time across the donor. In other words, throughout the shocking, the donor has no time to deviate greatly from its equilibrium shape under the Roche potential.

\subsection{Bound material}

We define bound material as having a Bernoulli parameter $\Be< 0 $. It is defined as
\begin{equation}
    \Be = \frac{1}{2} \left( \vel - \vel_{\mathrm{COM,bound}} \right)^{2} + \frac{\Gamma}{\Gamma - 1} \frac{ P }{ \rho } + \Phi, 
\end{equation}
where $\vel_{\mathrm{COM,bound}}$ is the center-of-mass of the bound material. While ideally one would iteratively find $\vel_{\mathrm{COM,bound}}$ so that $\Be$ is defined self-consistently \citep[e.g.][]{Guillochon2013,Prust2019}, we find that using the $\vel_{\mathrm{COM,bound}}$ from the previous timestep is sufficient to achieve fractional errors $< 10^{-5}$ in integrated quantities such as the total bound mass, $M_{\mathrm{bound}}$.

\subsection{Stopping condition}

Due to the kick from the SN ejecta, the donor moves along the $+x$-axis (to the right). 
In order to allow more time for the donor material to settle down, we keep the donor inside the simulation box by removing its center-of-mass momentum once it reaches the center of the box. This typically happens around $\approx 80\,t_{0}$ after explosion. We end the simulation once oscillations from the donor are nearly damped out (see the beginning of Section \ref{sec:fiducial}). 

\section{SN Ejecta modeling}
\label{sec:ejecta}

We adopt an SN ejecta profile motivated by the sub-$\mch$ bare CO core explosion models of \cite{Shen2018_subMch}, which have total masses of $0.9,1.0$ and $1.1\,\msun$, and C/O ratios of 50/50 and 70/30. 
Prior to encountering the donor, we assume homologous and adiabatic expansion. The adiabatic assumption is because the timescale prior to the ejecta-donor interaction is so short ($\lesssim 100$~s) that radioactive heating is of no consequence. 
Following \cite{Bauer2019}, we exclude ejecta with initial $v>$~20,000~$\kms$. We also account for the relative orbital motion by shifting the ejecta $y$-velocity. 

\begin{figure}[]
\centering
\fig{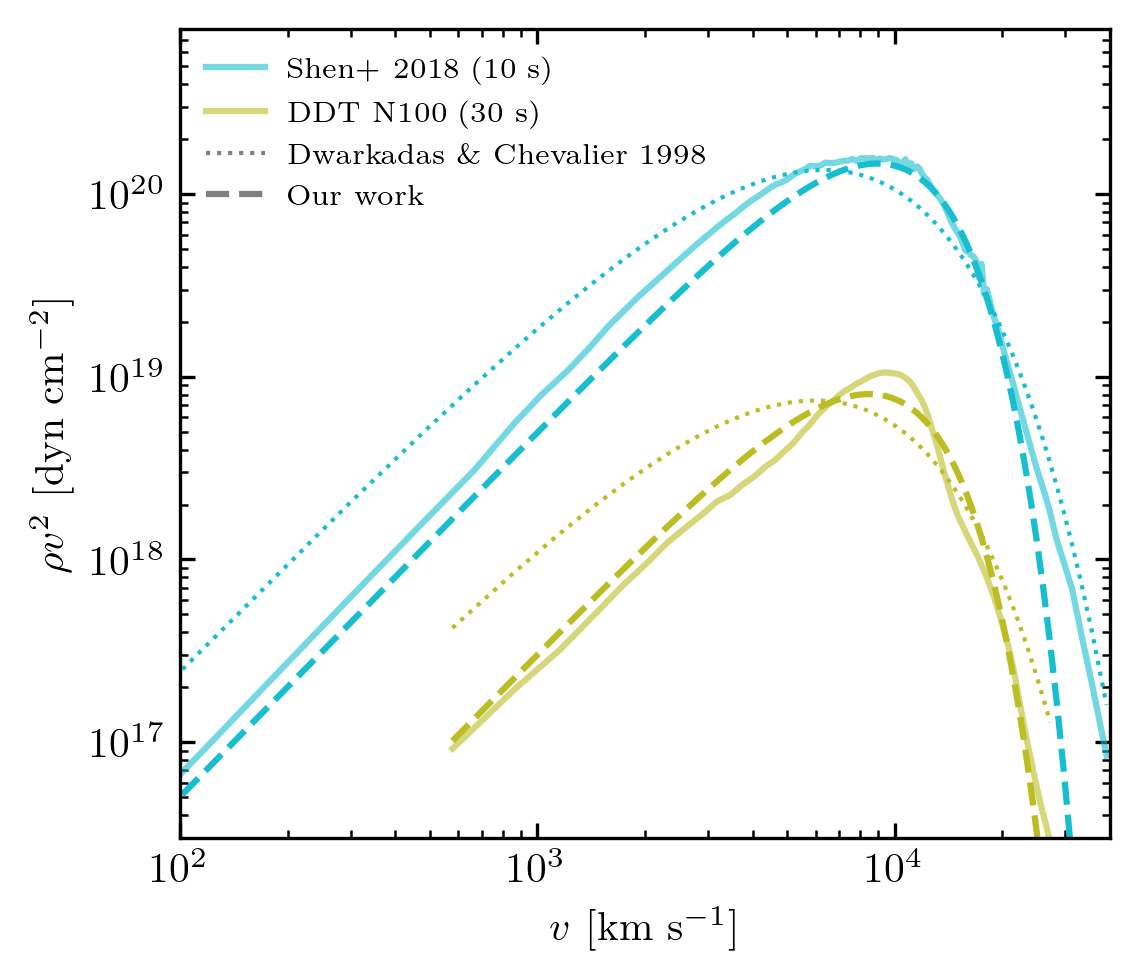}{ 0.5 \textwidth }{}
\caption{ 
Ejecta ram pressure. 
The $\Mej=1.0\,\msun$, C/O=50/50 model ($\Eej=1.2\times10^{51}\,\mathrm{erg}$, $t=10\,\mathrm{s}$) from \cite{Shen2018_subMch} is shown as a blue solid line. The Chandrasekhar-mass delayed-detonation model N100 from \cite{Seitenzahl2013} ($\Eej=1.5\times10^{51}\,\mathrm{erg}$, scaled to $t=30\,\mathrm{s}$) is shown as a yellow line. Our Gaussian ejecta profile and the exponential ejecta profile from \cite{Dwarkadas1988} are shown as dashed and dotted lines respectively, with the color corresponding to the \cite{Shen2018_subMch} or \cite{Seitenzahl2013} models. 
\label{fig:ejecta_profile}}
\end{figure}

While \cite{Bauer2019} adopted a broken power-law ejecta profile \citep[][]{Kasen2010}, and many other works have adopted an exponential ejecta profile \citep[][]{Dwarkadas1988}, here we introduce a Gaussian ejecta profile, defined by a total ejecta mass $\Mej$, total kinetic energy $\Eke$, and time $t$ since explosion, 
\begin{equation}
\label{eqn:rho}
    \rho( v, t)
    = 
    \left( \frac{ 3 }{ 4 \pi } \right)^{3/2} \frac{ \Mej^{5/2} }{ \Eke^{3/2}} \frac{ \exp \left[ - \left(v / v_{0} \right)^{2} \right] }{ t^{3} } ,
\end{equation}
where the ram pressure $\rho v^{2}$ peaks at a value
\begin{equation}
\label{eqn:v0}
    v_{0} 
    = 
    \left( \frac{ 4 }{ 3 } \frac{ \Eke }{ \Mej } \right)^{1/2} .
\end{equation}
The Gaussian ejecta profile shows better agreement across all total ejecta masses with the sub-$\mch$ CO core explosion models of \cite{Shen2018_subMch}, than the broken power-law and exponential forms. 
A comparison among the $1.0\,\msun$, C/O=50/50 model from \cite{Shen2018_subMch}, our Gaussian ejecta profile and the exponential ejecta profile \citep[][]{Dwarkadas1988} is shown in Figure \ref{fig:ejecta_profile}. 
Our Gaussian ejecta profile also yields a constant density core shown by the models \citep[][]{Shen2018_subMch}, and agrees well in the shape and location of the peak of the ram pressure $\rho v^{2}$. For $v < $\,20,000\,$\kms$, the agreement in $\rho$ and hence $\rho v^{2}$ is within 20\%, which is further reduced to 10\% near the peak of the ram pressure. We also show the Chandrasekhar-mass delayed-detonation model N100 from \cite{Seitenzahl2013}, scaled by a factor $(100/30)^{3}$. Despite the different explosion physics, our Gaussian ejecta profile also shows better agreement than the exponential profile. As our prime goal in this work is in assessing the impact on the donor from the ejecta, we place a premium on the agreement of the $\rho v^{2}$ profile. 

As noted earlier, at the time of injection into the {\athena} grid, the fluid has expanded adiabatically and the flow is approaching homology. Hence we are not so concerned about the explicit choice of the ejecta internal energy. 
\cite{Bauer2019} set the internal energy density of the ejecta to be $\left( 0.08 a / r_{a} \right) \rho v^{2}$, where $r_{a}$ is the distance from the accretor/explosion center, such that after undergoing homologous expansion, the internal energy density was about 7-9\% of the kinetic energy density when the ejecta reaches the donor. 
Instead, we take advantage of the fact that fluid elements in an adiabatic flow with $\Gamma={5/3}$ maintain a constant $P/\rho^{5/3}$ until they are shocked by the donor collision. 
Rather than fitting a profile of $P/\rho^{5/3}$ from the explosion models of \cite{Shen2018_subMch}, we adopt a fixed value of $\langle P/\rho^{5/3} \rangle = 0.59 \times 10^{14} $ in cgs units, which is the mass-averaged value of the $\Mej=1.0\,\msun$, C/O=30/70 model (averaged over $v <$~10,000~$\kms$). The pressure is then given by $P(v,t) = \langle P/\rho^{5/3} \rangle \left( \rho(v,t) \right)^{5/3}$, where $\rho(v,t)$ is given by Eqn \ref{eqn:rho}. Application to other models shows that the internal energy density agreement is within $50\%$ for $v \lesssim$~10,000~$\kms$. However, in all cases the internal energy density remains $\lesssim 10\%$ of the kinetic energy density. 
Hence, none of these choices have dramatic impact on the ejecta-donor interaction. 
In this paper, we choose explosion kinetic energies $E_{\rm KE}$ of $1.5,\,1.2,\,0.8$~\&~$0.5\times 10^{51}\,\mathrm{erg}$.

\section{Outcomes for a helium WD donor with $E_{\rm KE}=1.2\times10^{51}\,\mathrm{erg}$}
\label{sec:fiducial}

\begin{figure*}[t!]
\centering
\fig{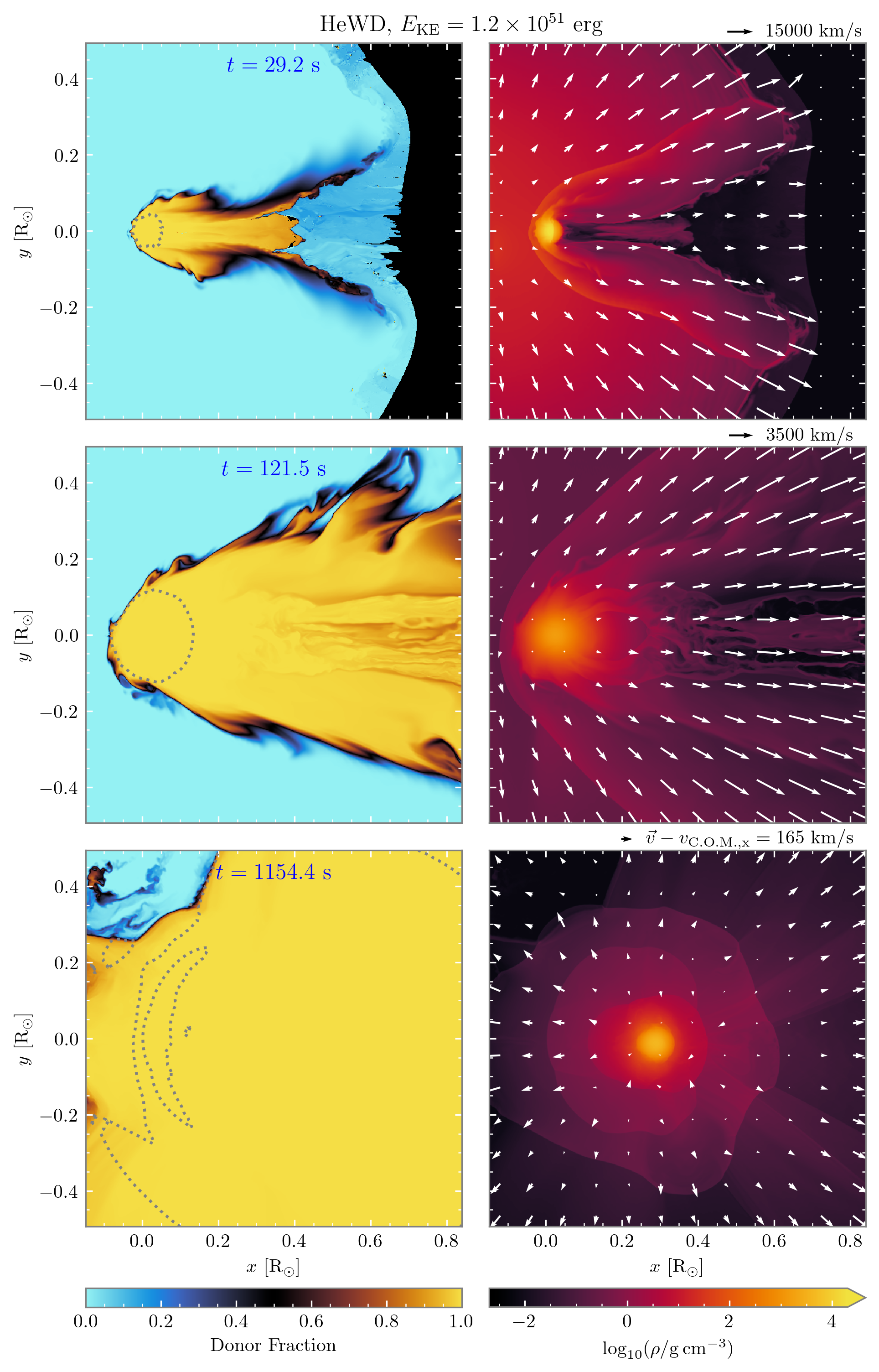}{ 0.78 \textwidth }{}
\caption{ 
Fraction of donor material (left column) and total density (right column) at the mid-plane of the simulation, taken at different snapshots, for the HeWD model with $E_{\rm KE} = 1.2 \times 10^{51}\,\rm{erg}$. In the right column, white arrows indicate the velocity field, and we give the arrow legend in the upper right corner. In the bottom snapshot, the velocity field is relative to the $x$-velocity of the donor remnant center-of-mass (219 km/s). Time since the injection of SN ejecta is indicated in the left column for each snapshot. Grey dotted contours in the left column shows bound material with $\mathrm{Be} < -0.01 P^{i}_{\rm c}/\rho^{i}_{\rm c}$, and illustrates that most material remaining in the grid at the end of the simulation is bound.
\label{fig:TimeSequence}}
\end{figure*}

We show three different snapshots of the HeWD, $\Eke=1.2\times10^{51}\,\mathrm{erg}$ model in Figure \ref{fig:TimeSequence}. 
As others have shown, the bow shock extends to an angle that is roughly 
 twice that subtended by the star, $\theta_{\rm wake} \approx  \arctan \left( 2 \Rhe / a \right) = 25^{\circ}$ \citep[e.g.,][]{Kasen2010}, implying that $\approx10\%$ of the ejecta is modified. 
The ejecta flows around the donor, entraining donor material mostly from the upstream side (facing the explosion center), and leaving behind a low-density conical hole in the wake behind the donor. 
Meanwhile, the donor is compressed by the ejecta and a shock propagates through it, visible in the top panel of Figure \ref{fig:TimeSequence}. 

The shock eventually breaks out from the downstream side of the donor (facing away from the explosion center). Unbound, expanding donor material from the shock breakout quickly fills up the conical hole behind the donor. The shocked donor then expands and moves to the right due to momentum transfer from the ejecta. This is visible in the middle panels of Figure \ref{fig:TimeSequence}.

As the donor adjusts to a new hydrostatic equilibrium, it contracts and expands periodically with decreasing amplitude. With each contraction, a shock wave is sent radially outwards. We find that the passing of the first shock wave greatly increases the entropy of the outermost donor material, whereas subsequent shocks do little to raise the entropy. The shock waves sent by the oscillating donor are seen as nearly concentric rings of enhanced density in the bottom panels of Figure \ref{fig:TimeSequence}. This snapshot is near the end of the simulation, and at this moment almost all material in the grid, consisting mostly of donor material, is bound. The flow is nearly radial around the donor in the center-of-mass frame. Material flows slightly slower at the downstream side due to the slightly higher density there. 

\subsection{Origin of Unbound Donor Mass}
\label{sec:scalar_results}

\begin{figure}
\centering
\fig{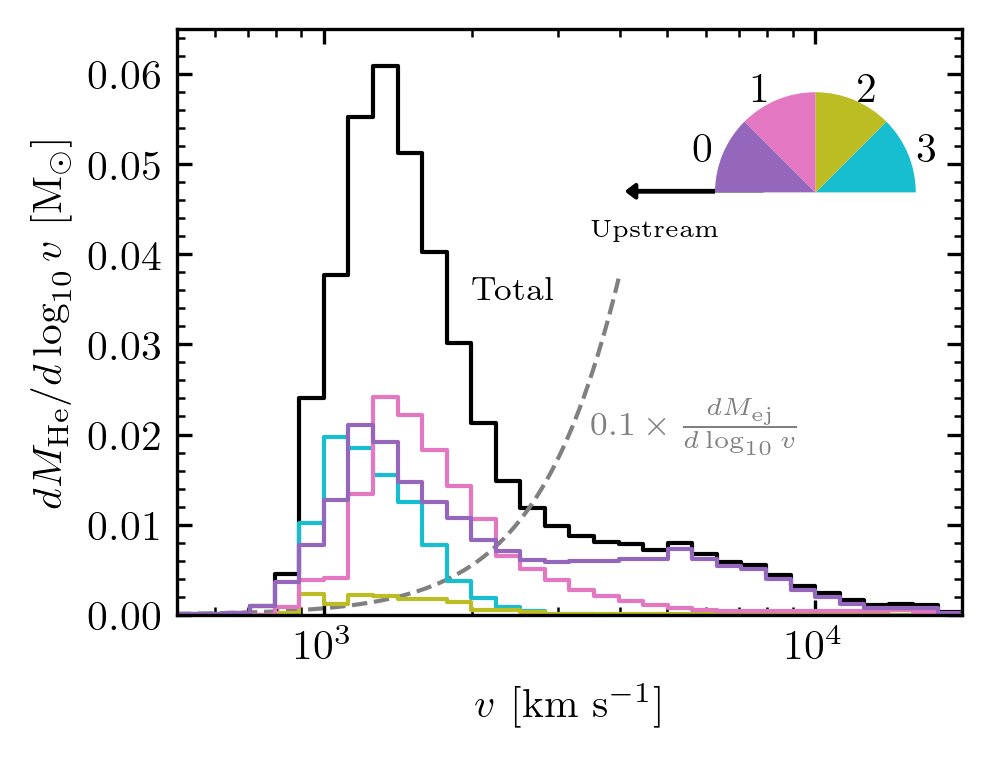}{ 0.5 \textwidth }{}
\caption{ 
Mass loss velocity distribution from different sectors of the HeWD donor for $\Eke=1.2\times10^{51}$~erg. The top right inset shows the different sectors of the donor relative to the explosion center, each with a different color corresponding to the curves in the mass loss distribution. The black curve represents the total mass loss distribution from all four sectors. The grey dashed line shows the ejecta velocity distribution, scaled by fraction of solid angle intercepted by the donor ($\approx 10~\%$). 
\label{fig:ML_distribution}}
\end{figure}

To clarify where most of the unbound He originates, we reran the HeWD, $\Eke=1.2\times10^{51}\,\mathrm{erg}$ model but with four different passive scalars for different sectors of the donor. 
As displayed in Figure \ref{fig:ML_distribution}, scalar 0 faces the explosion center whereas scalar 3 faces the backside. 
We record the velocity of the unbound donor material leaving the simulation box originating from each sector, and show the distribution in Figure \ref{fig:ML_distribution} after accounting for motion relative to the binary center-of-mass. The total unbound donor mass loss is $\approx 0.013~\msun$, which is only $\approx 10\%$ of the initial donor mass. 

There is a high-velocity tail at $v \geqslant$~5,000~$\kms$ dominated by mass loss from the upstream hemisphere, with higher contributions from sector 0. This is not surprising because the upstream hemisphere is subject to strong momentum transfer from the SN ejecta. 

Mass loss from the upstream hemisphere (sectors 0 and 1) increases towards lower velocities and peaks at $\approx$~1,300\,$\kms$. Around the same velocity, we also find significant mass loss from sector 3, the downstream sector facing directly away from the explosion center. The mass loss there is not a consequence of direct momentum transfer from the SN ejecta, but rather due to ablation from shock breakout down the density gradient. In all velocity ranges, sector 2 contributes to little mass loss, whereas sectors 0, 1 and 3 collectively contribute to a significant peak in mass loss at  $v\approx$~1,300\,$\kms$. We note that even without accounting for orbital motion of the donor, the mass loss peak would still peak at $\approx$~900\,$\kms$. 

We compare the donor mass loss velocity distribution to that of the ejecta scaled by 0.1, approximately the fraction of solid angle intercepted by the donor. Assuming that most of the donor mass loss is constrained within $ \approx 30^{\circ}$ downstream from the explosion center (see next section), then we predict that donor material dominates over ejecta at $v \lesssim\,$2,500\,$\kms$ at these angles in the wake. Our prediction that donor mass loss dominates at low velocities is robust -- even without scaling down $ d M_{\rm ej} / d \log_{10} v $ by $0.1$, the donor material dominates at $v \lesssim\,$1,800\,$\kms$. We also note that since the donor mass loss lies mostly at $\lesssim$~10,000~$\kms$, the conical hole in the ejecta cannot be filled with donor material at high velocities.

\subsection{Angular distributions of donor material and ejecta}

\begin{figure}
\centering
\fig{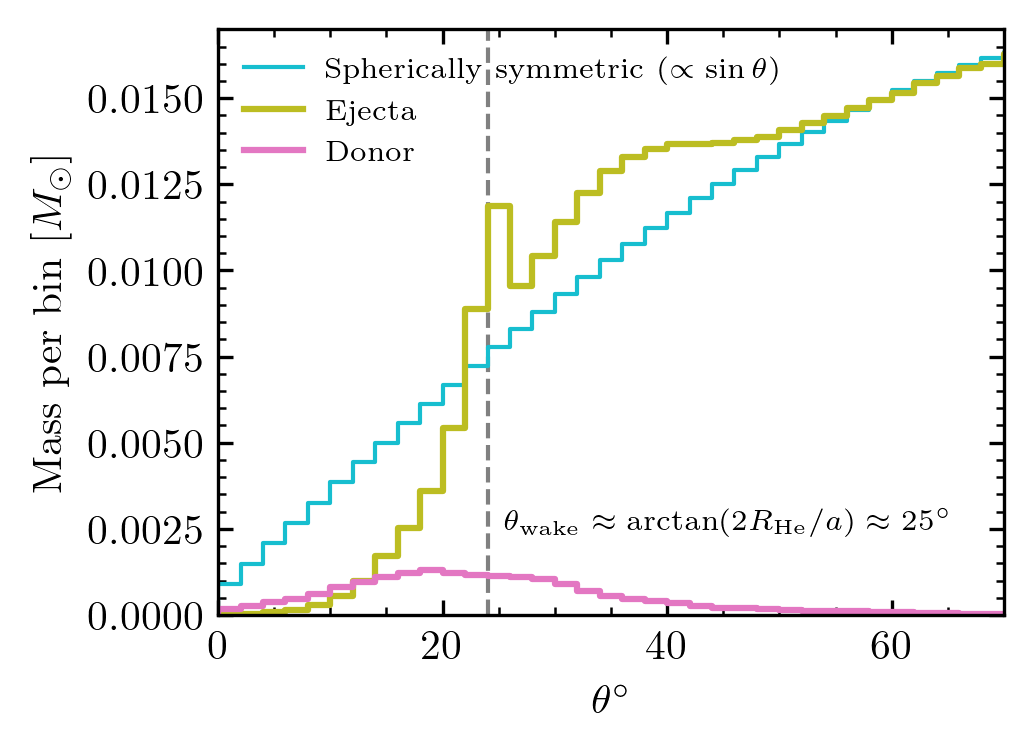}{ 0.5 \textwidth }{}
\caption{ 
Angular distribution of donor mass loss (pink) and ejecta (yellow) for $\Eke=1.2\times10^{51}$~erg. Blue line shows the expected ejecta distribution \textit{if} it were spherically symmetric. Dashed grey line shows $\theta_{\rm wake} \approx \arctan(2 \Rhe / a) \approx 25^{\circ}$, slightly beyond which a bow shock is formed.
\label{fig:angular_distribution}}
\end{figure}

Figure \ref{fig:angular_distribution} shows the angular distribution of unbound mass leaving the simulation box. 
We define the angle $\theta$ relative to the explosion center, such that $\theta = 0^{\circ}$ points downstream. 
For a spherically symmetric flow, the mass per solid angle is $1\,M_{\odot}/(4 \pi)$, and so the mass between angles $\theta$ and $\theta + d\theta$ is given by $1\,M_{\odot}/(4 \pi) \times 2 \pi \times [ \cos\theta - \cos(\theta + d\theta) ] \propto \sin \theta d\theta $, i.e. at larger angles, the mass per $\theta$ bin increases due to a larger solid angle. This is observed for the ejecta at $\theta \gtrsim 50^{\circ}$. However, for $\theta \lesssim 50^{\circ}$, the flow of the ejecta deviates from spherical symmetry because it has to divert around the donor. In particular, $\theta \lesssim \theta_{\rm wake} \approx 25^{\circ}$ is relatively devoid of ejecta, and at $\theta \approx 30^{\circ}$, approximately where a bow shock is formed, there is a bump in ejecta. 
On the other hand, we find a peak in donor mass loss at $\theta \approx 20^{\circ}$ extending to $\approx 40^{\circ}$. Donor material dominates over the ejecta at $\theta \lesssim 15^{\circ}$. 
However, our results should be taken with a grain of salt since we adopt a constant $\Gamma = 5/3$, but the post-shock ejecta should be radiation-pressure dominated $(\Gamma = 4/3)$. 

\subsection{Origin of bound ejecta material}

To identify the composition of the ejecta material that remains bound to the donor, 
we reran the HeWD, $\Eke = 1.2 \times 10^{51}$~erg model with two passive scalars for the ejecta, distinguished by whether their incoming velocity is below $10$,$000$~$\kms$. This choice because the $\Mej=1.0\,\msun$ model in \cite{Shen2018_subMch} is dominated by iron group elements (IGE) below $10$,$000$~$\kms$. We find that $\approx 85\%$ of the bound ejecta material originates from $v \lesssim 10$,000~$\kms$, thus showing that most of it is IGE in composition.

\section{He WD donor with other $\Eke$}
\label{sec:other_Eke}

\begin{figure*}
\centering
\fig{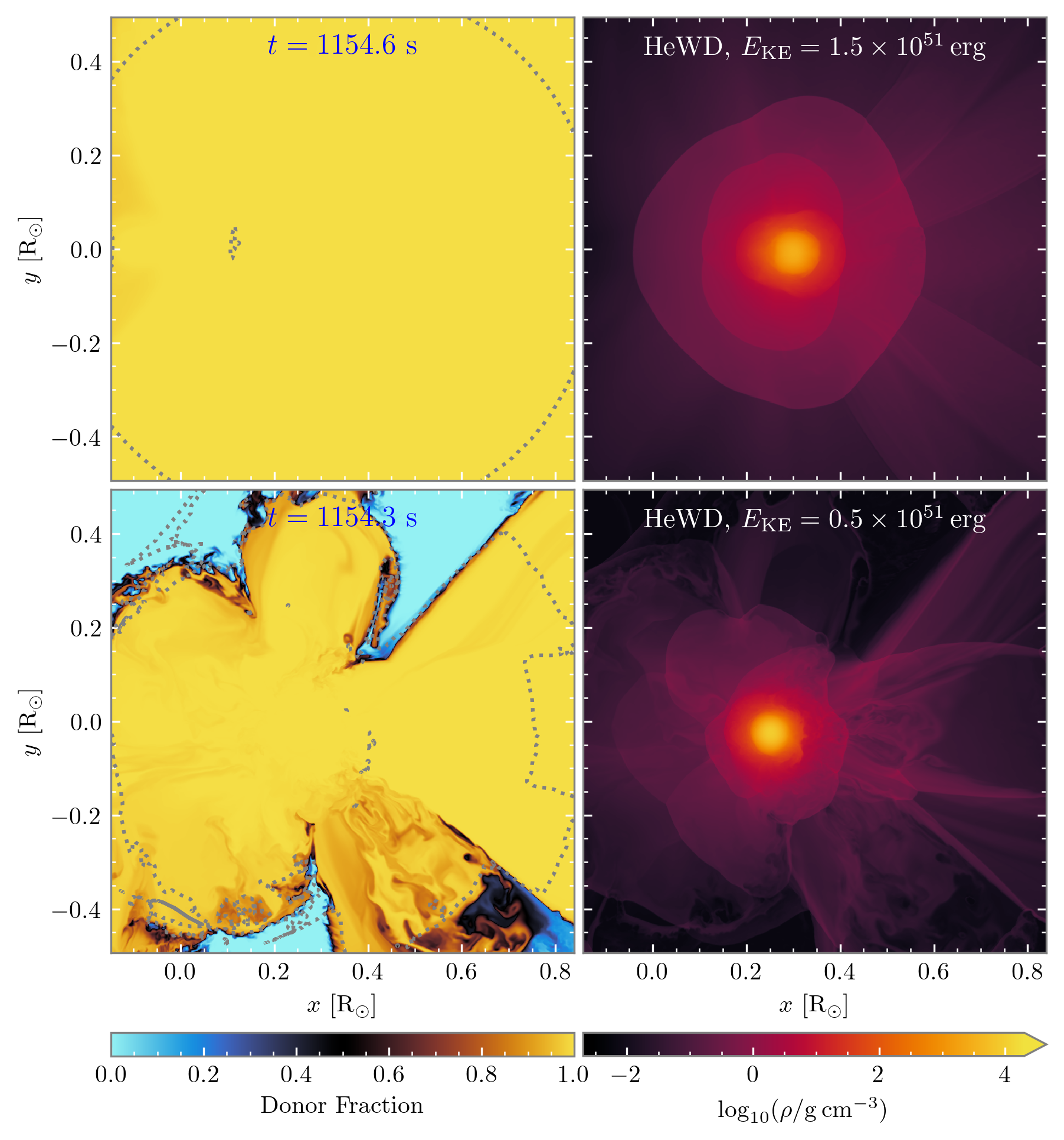}{ 0.75 \textwidth }{}
\caption{ 
Same as Figure \ref{fig:TimeSequence}, but at $t=1154$~s and with $\Eke=1.5$ (top row) \& $0.5\times10^{51}$~erg (bottom row). 
\label{fig:Eexp_snapshot}}
\end{figure*}

We now discuss the results of the HeWD simulations with other $\Eke$. 
Figure \ref{fig:Eexp_snapshot} shows snapshots at $t=1155$~s for $\Eke=1.5$~\&~$0.5\times10^{51}$~erg. As expected, a higher $\Eke$ deposits more entropy into the HeWD donor, causing the donor remnant to be much more extended, so much that the entire grid is filled with bound donor material. In contrast, the low $\Eke$ simulation shows a more centrally concentrated donor remnant, with some bound ejecta in the grid. 
We can also see that the donor has moved further to the right for $\Eke=1.5\times10^{51}$~erg, which is a result of the greater momentum transfer from the ejecta. 

\begin{figure}
\centering
\fig{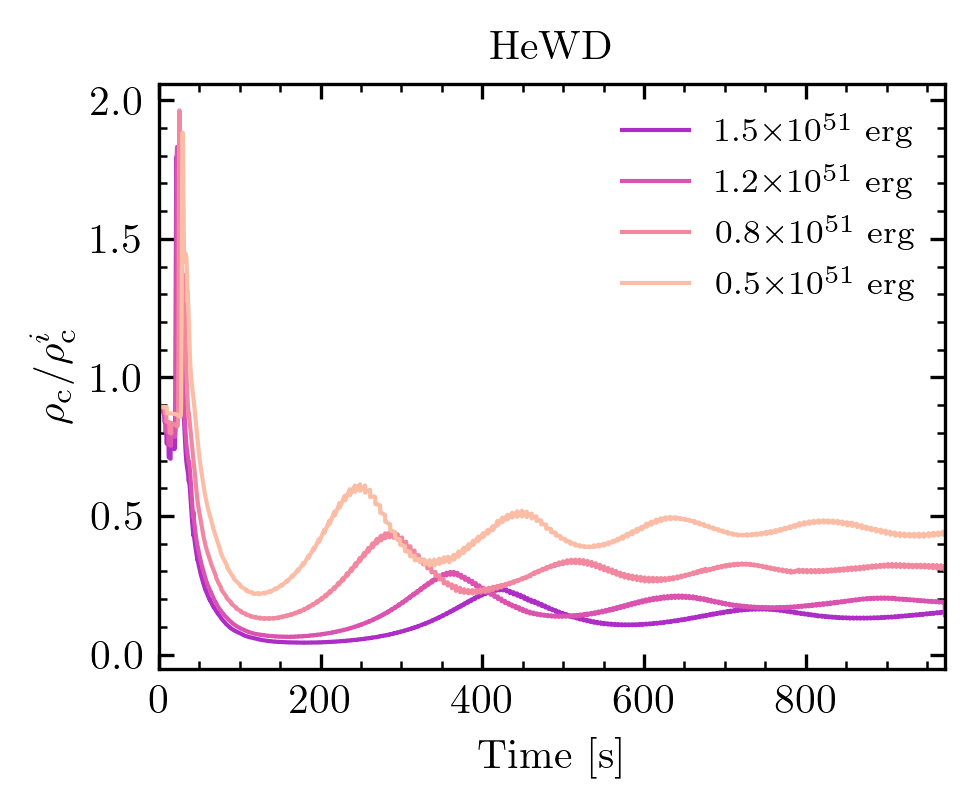}{ 0.5 \textwidth }{}
\caption{ 
Central density of the HeWD donor for different $\Eke$. 
\label{fig:HeWD_rhoc}}
\end{figure}

Figure \ref{fig:HeWD_rhoc} shows the density at the center-of-mass of the donor for different $\Eke$. Right before the ejecta is injected, the central density is slightly lower than at the beginning of the simulation, due to relaxation of the donor in the co-rotating frame. As a shock propagates through the donor, the central density jumps by roughly a factor of 2, and falls subsequently as the donor expands. For the rest of the simulation, the central density oscillates around some equilibrium value as the donor rings. With higher $\Eke$, the donor has a lower $\rhoc$ since it is more loosely bound, and a longer oscillation period. 
The oscillation periods shown by the models roughly agree with the fundamental mode pulsation period of a polytrope \citep[][]{Cox1980}, with a polytropic index chosen to approximately match the density profile. 

Table \ref{tab:athena_results} shows the final change in bound helium mass $\Delta M^{f}_{\rm He}$, bound ejecta mass $M^{f}_{\rm ej}$, final central density of the donor $\rhoc^{f}$ in units of the initial central density $\rhoc^{i}$, its change in central specific entropy $\Delta \Sc$, and the kick velocity in $x$- and $y$-directions, $v_{\rm kick,x}$ \& $v_{\rm kick,y}$, for different $\Eke$. 
We estimate the change in donor central specific entropy as follows:
\begin{equation}
\label{eqn:entropy}
    \Delta \Sc = \frac{3}{2} \frac{ \kB }{ \mu m_{\rm p} } \ln \left[ \frac{ P_{\rm c}^{f} }{ P_{\rm c}^{i} } \left( \frac{ \rho_{\rm c}^{i} }{ \rho_{\rm c}^{f} } \right)^{5/3} \right] ,
\end{equation}
where $\mu\approx 4/3$ is the mean molecular weight, and $P^{f}_{\rm c}$ is the final central pressure.

With higher $\Eke$, mass loss from the donor increases, with a roughly linear scaling in agreement with \cite{Pakmor2008}. In general, we find that the donor loses $\approx 0.01-0.025\,\msun$. 
Meanwhile, $\sim 10^{-7}-10^{-4}\,\msun$ of SN ejecta stays bound to the donor surface, which we consider as lower limits, since our finite box sizes do not allow us to track all ejecta material. 
A higher $\Eke$ causes greater shock deposition into the donor and hence higher $\Delta \Sc$. In turn, the donor experiences greater expansion, and hence lower $\rhoc^{f}$.

With higher $\Eke$, the donor also receives a greater kick velocity. 
We find that the $x$-momentum intercepted by the donor is about $\approx 18 - 24$\% of that expected, 
in the range of $\approx 1/3$ as predicted by \cite{Hirai2018} and \cite{Bauer2019}, yielding $v_{\rm kick,x}\approx 170 - 230\,\kms$. The donor also receives a kick in the negative $y$-direction $\approx$ few $\kms$ due to the orbital motion. However, the final spatial velocity is only $\approx 2\%$ higher than $v_{\rm orb}$, since the kick direction and the donor orbital motion are mostly perpendicular to each other \citep[see also Figure 15 of][]{ElBadry2023}.

\cite{Papish2015} and \cite{Tanikawa2019} showed that a $0.4-0.45\,\msun$ He WD donor can undergo a detonation after being shock-heated by the supernova ejecta. Although nuclear reactions are not accounted for in our {\athena} simulations, we check for this possibility in our post-processing. For each cell, we solve for the temperature given the density and pressure, using the {\mesa} equation-of-state module, and calculate the local heating timescale $c_{\rm p} T / \epsilon_{\rm nuc}$ using the {\mesa} nuclear net module, where $c_{\rm p}$ is the specific heat capacity at constant pressure and $\epsilon_{\rm nuc}$ is the specific nuclear energy generation rate. Although some cells, particularly at the upstream side of the donor, may momentarily get shock-heated to above $10^{8}$~K, we find that the local heating timescale is always longer than $10^{4}$~s, which is still longer than our simulation time. This suggests that helium detonation does not occur in our HeWD simulations. This differs from \cite{Papish2015} and \cite{Tanikawa2019} likely because our He WD donor is of lower mass ($0.13\,\msun$) and therefore density.

\begin{deluxetable*}{ccccccccccc}
\tablenum{2}
\label{tab:athena_results}
\tablecaption{Final results}
\tablehead{
\colhead{Donor Model} & 
\colhead{$\Eke$} &
\colhead{$\Delta M^{f}_{\rm He}$} &
\colhead{$M^{f}_{\rm ej}$} &
\colhead{$\rho^{f}_{c}$} &
\colhead{$\Delta s_{\rm c}$} &
\colhead{$v_{\rm{kick},x}$} &
\colhead{$v_{\rm{kick},y}$} &
\colhead{$v_{\rm total}$} &
\\
\colhead{} &
\colhead{[$10^{51}\,\mathrm{erg}$]} &
\colhead{[$\msun$]} &
\colhead{[$\msun$]} &
\colhead{[$\rho^{i}_{c}$]} &
\colhead{[$\kB/\mu m_{\rm p}$]}&
\colhead{[$\kms$]} &
\colhead{[$\kms$]} &
\colhead{[$\kms$]} 
}
\tablewidth{100pt}
\startdata
HeWD & $1.5$ & 0.025 & $10^{-7}$ & 0.14 & 0.29 & 227 & -5.7 & 975 \\
     & $1.2$ & 0.020 & $10^{-5}$ & 0.19 & 0.26 & 220 & -10 & 969 \\
     & $0.8$ & 0.015 & $10^{-4}$ & 0.32 & 0.15 & 195 & -8.7 & 965 \\
     & $0.5$ & 0.010 & $10^{-4}$ & 0.46 & 0.15 & 166 & -16 & 952 \\
\hline
HeStar1 & $1.5$ & 0.010 & $10^{-4}$ & 0.80 & 0.04  & 121 & -1.5 & 726 \\
        & $1.2$ & 0.008 & $10^{-4}$ & 0.83 & 0.03 & 112 & -2.3 & 724 \\
        & $0.8$ & 0.005 & $10^{-3}$ & 0.88 & 0.02 & 94 & -3.4 & 720 \\
\hline
HeStar2 & $1.5$ & 0.016 & $10^{-4}$ & 0.48 & 0.10 & 247 & -4.9 & 1017 \\
        & $1.2$ & 0.012 & $10^{-4}$ & 0.55 & 0.08 & 238 & -6.1 & 1013 \\
        & $0.8$ & 0.009 & $10^{-4}$ & 0.67 & 0.07 & 210 & -9.0 & 1004 \\
\enddata
\tablecomments{ Here $\Eke$ is the total SN kinetic energy, $\Delta M^{f}_{\rm He}$ is the mass loss from the donor, $M^{f}_{\rm ej}$ is a rough estimate of the bound SN ejecta mass, $\rho^{f}_{\rm c}$ is the final central density of the donor, $\Delta s_{\rm c}$ is the change in central specific entropy of the donor as estimated by equation \ref{eqn:entropy}, $v_{\rm kick,x}$ and $v_{\rm kick,y}$ are the kick velocities received by the donor in the $x$ and $y$-directions, and $v_{\rm total}$ is the final ejection velocity of the donor after accounting for the kick and its orbital motion. }
\end{deluxetable*}

\section{Helium star models}
\label{sec:HeStar}

We revisited the He star models from \cite{Bauer2019} for comparison and to explore the effects of different numerical treatments. These are summarized in Appendix \ref{appendix_A}. 
We now discuss simulations with the HeStar1 and HeStar2 donor models using the fiducial setup described in Section \ref{sec:numerics}. 
The results are summarized in Table \ref{tab:athena_results}. 

Based on the discussion in \cite{Bauer2019}, we expect that the ratio between the ejecta pressure at the donor location and the donor central pressure is given by:
\begin{equation}
    \frac{ P_{\rm ej} }{ P_{\rm c} } \approx \frac{7 \times 10^{-3}}{ q^{2/3} (1+q)^{7/3} } \left( \frac{v_{\mathrm{ej}}}{v_{\mathrm{orb}}}\right)^{2} ,
\end{equation}
where $q = \mhe / \mwd$ is the mass ratio. Although donors in wider orbits tend to suffer more entropy impact due to the $v_{\rm orb}$ scaling, after also accounting for the mass ratio $q$ we expect the HeWD donor to be the most impacted. Indeed, comparing the results in Table \ref{tab:athena_results} shows that the HeWD donor gains much more entropy than the He star models, and as a result of the expansion experiences a steeper drop in $\rhoc$. 
In general, the He star donors lose $\approx 0.01\,\msun$, slightly less than the HeWD model. 
The velocity distribution peaks at $v \approx $~1,100 \& 1,600\,$\kms$ for HeStar1 and HeStar2 respectively. 
Also, $\approx 10^{-4}\,\msun$ of SN ejecta stays bound to the He star donors. 
Like the He WD models, we find that a helium detonation is unlikely for the He stars.

\section{Post-explosion evolution in MESA}
\label{sec:postexp}

After the oscillations in the shocked donor are damped, we continue its subsequent evolution on much longer timescales using the stellar evolution code {\mesa} version 24.03.1. 
We first adjust the {\mesa} donor model to the appropriate post-explosion mass. 
Since {\athena} does not evolve $T$ as a base variable, we obtain spherical averages of $P$ and $\rho$ from {\athena}, and then iterate to find the corresponding $T$ profile using {\mesa}'s {\tt eos} module. 
We then adjust the {\mesa} model to this $T-\rho$ profile using {\mesa}'s {\tt relax\_initial\_entropy} routine.
We do not take into account pollution by the SN ejecta, and adopt solar metallicty ($Z=0.014$) for the opacity. We also do not consider rotation. Convection is treated with {\mesa}'s time-dependent convection formulation \citep[][]{MESAVI} and a mixing length parameter $\alpha_{\rm MLT} = 1.89$. 

\begin{figure}
\centering
\fig{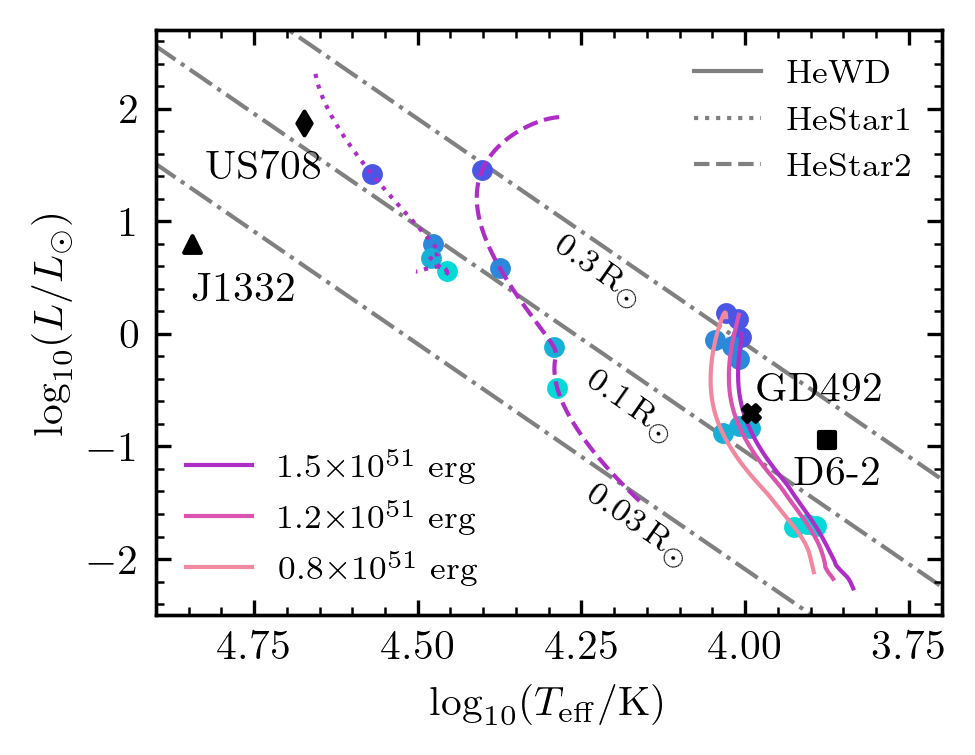}{ 0.5 \textwidth }{}
\caption{ 
Post-explosion donor evolution on the Hertzsprung-Russell diagram. For the HeWD donor (solid), we show the $\Eke = 1.5,1.2\,\&\,0.8\times10^{51}$~erg models. For the HeStar1 (dotted) and HeStar2 (dashed) donors, we only show the $\Eke=1.2\times10^{51}$~erg models. All evolutionary tracks start from an age of $10^{3}$~yrs and end at $10^{8}$~yrs. The colored circles label ages of $10^{4}$, $10^{5}$, $10^{6}$ \& $10^{7}$ yrs. For comparison, we show lines of constant radius (dot-dashed lines), and the runaway stars D6-2 (square), J1332 (triangle) and US 708 (diamond). 
\label{fig:HRD}}
\end{figure}

We show the post-explosion evolution of the donor on the Hertzsprung-Russell diagram (HRD) in Figure \ref{fig:HRD}. 
The entropy deposited by the ejecta increases toward the outer layers of the donor. As a result, all donor models initially show a temperature inversion in their outer layers. 
The HeStar1 and HeStar2 models brighten and expand to $\approx 10^{3}\,\lsun$ and $10^{2}\,\lsun$ within the first $\approx 10$ and $10^{2}$ yrs, respectively, while the HeWD model does not. 
Similar brightening is shown by other works, for main sequence donors \citep[][]{Pan2012b,Liu2021b,Rau2022} and He star models \citep[][]{Pan2013,Bauer2019,Liu2021a,Liu2022,Zeng2022}. 
However, we do not show this short-lived phase, corresponding to the first $10^{3}$ yrs, in Figure \ref{fig:HRD}. As noted by others, objects this bright are of interest to search for in young supernova remnants \citep[e.g.,][]{Shields2022}.

After $\approx 10^{4}-10^{5}$ yrs, roughly corresponding to the local heat diffusion timescale of peak heating \citep[][]{Bauer2019,Zhang2019}, the donor envelope loses entropy and the temperature inversion vanishes. The donor dims and contracts. 
The HeWD and HeStar2 models keep on cooling, but the HeStar1 model eventually resumes core He burning after $\approx 10^{6}$ yrs, as its mass is above the limit for He burning ($\approx 0.3\,\msun$). It continues core He burning for the next $\approx 10^{8}$ yrs.

We now compare our models to several runaway stars. The hypervelocity star D6-2 has reliable astrometry from Gaia EDR3 and photometry, allowing a fit to the spectral energy distribution that yields $T_{\rm eff} = 7500 \pm 100$~K and $R = 0.20 \pm 0.01\,\rsun$ \citep[][]{Chandra2022}. Its trajectory traces back to a known supernova remnant, allowing derivation of a kinetmatic age of $\approx 10^{5}$~yrs \citep[][]{Shen2018_D6}. 
The HRD evolution of our HeWD model, whose pre-explosion $v_{\rm orb}$ agrees well with the ejection velocity of D6-2 \citep[][]{Wong2023}, agrees decently with the location of D6-2. However, it is somewhat over-luminous compared to D6-2 at an age of $10^{5}$ yrs.

J1332-3541 is among the 4 newly discovered hypervelocity stars \citep[][]{ElBadry2023}, and has an uncertain radius $R = 0.017 ^{+0.013}_{-0.007} \, \rsun$ due to its uncertain distance. We adopt a temperature of $70,000$~K \citep[see also][]{Werner2024}. 
US 708 is a He-rich subdwarf O star, with an inferred disk-crossing $\approx 10^{7}$~yrs ago. We adopt $T_{\rm eff} = 47200$ and $\log g = 5.69$ to calculate its luminosity assuming a mass of $0.3\,\msun$. 
The He star models have $T_{\rm eff}$ between D6-2, and J1332 and US 708. 
At an age of $\approx 10^{7}$~yrs, the HeStar1 model settles near the He main sequence, but appears too dim and red compared to US 708. 
In agreement with \cite{Bauer2019}, we find that if US 708 is currently a He-burning star, its mass must be higher than that of HeStar1 ($\approx 0.35\,\msun$). However, \cite{Neunteufel2020} inferred a mass of $0.34-0.37\,\msun$ for US 708, not much different than our HeStar1 model. Alternatively, it could still be undergoing thermal relaxation if the local thermal diffusion timescale at the location of peak heating is long. 

Interestingly, the HeWD trajectories pass near GD 492 (LP 40-365), which has an estimated $T_{\rm eff} = 9800\pm 300$~K and $L = 0.20 \pm 0.04\,\lsun$ \citep[][]{Raddi2019}. This star represents a class of runaway objects mostly with ejection velocities $\approx 600 - 700\,\kms$ \citep[][]{ElBadry2023} and Ne-dominated atmospheres, believed to be the surviving \textit{accretor} after incomplete burning fails to unbind the star \citep[][]{Raddi2019}. The estimated disk-crossing time for most of these objects is $\approx 10^{6}$~yrs, coinciding with the age at which the HeWD trajectories intersect GD 462 on the HR diagram.

We note that pollution by SN ejecta can greatly enhance the surface opacity. We find that adopting a solar-scaled metallicity of $Z=0.03$ shifts the HeWD, $\Eke=1.2\times10^{51}$~erg model redwards by $\approx$1,000$-$2,000~K, and a $Z=0.05$ model agrees well with the present-day location of D6-2 at an age of $\approx 10^{6}$~yrs \citep[though D6-2 has an inferred age of $\approx 10^{5}$ yrs;][]{Shen2018_D6}. Given a surface convection zone mass of $\approx 0.01\,\msun$ from the models, these $Z$ values require only a few $10^{-4}\,\msun$ of ejecta be deposited and stay on the donor surface, which is within the range of bound SN ejecta mass in our hydrodynamic models (see Table \ref{tab:athena_results}). 

D6-2 has a photometric period of $15.4$~hrs, inferred to be its rotation period \citep[][]{Chandra2022}. 
Although our models do not account for rotation of the donor, we can estimate the rotation rate. We assume that the pre-explosion donor is tidally synchronized to its orbit \citep[e.g.,][]{Fuller2012}, and uniformly rotating. 
We also assume that the donor conserves its total angular momentum of the remaining bound mass immediately after explosion, $J_{\rm i}$. 
In the limit of solid-body rotation, we can calculate the total moment of inertia $I_{\rm f} = \frac{2}{3} \int^{M_{\rm f}}_{0} r^{2} \,\mathrm{d}m $, and hence the rotation period $P_{\rm f} = 2 \pi I_{\rm i} / J_{\rm f} $. For the HeWD, $\Eke = 1.2 \times 10^{51}$~erg model, this implies a rigid rotation period of $\approx 1.7$~hrs at $10^{5}$~yrs, which is too rapid compared to D6-2. 
In the limit of no angular momentum transport, each fluid element retains its specific angular momentum $j(r) = i(r) \omega(r)$, where $i(r) = \frac{2}{3} r^{2}$ is the specific moment of inertia. 
The final surface rotation period is then $P_{\rm f} = P_{\rm i} \left(  R_{\rm f} / R_{\rm i} \right)^{2} $, where $R_{\rm f}$ is the final donor surface radius, and $R_{\rm i}$ is the corresponding pre-explosion radius of these outermost mass element. At an age of $10^{5}$, we find a rotation period of $18.5$~hr, in much better agreement with the photometric period of D6-2. 
This comparison suggests that D6-2 has not had sufficient time for interior angular momentum transport, or that it experienced angular momentum loss during the post-explosion evolution \citep[][]{Chandra2022}. 
As the donor remnant further contracts, its rotation period decreases. For example, its rotation period would be 0.8~hrs for rigid-body rotation at $10^{6}$~yrs, and correspondingly 4.1~hrs with no angular momentum transport. 

Similarly, the He star models start off slowly rotating but spin up as they contract. At $10^{5}$~yrs, the HeStar1 model has a rotational period of 0.3~hrs (1.2~hrs) for solid-body rotation (no angular momentum transport), which decreases to 0.2~hrs (0.9~hrs) at $10^{7}$~yrs. These values agree reasonably with \cite{Bauer2019}. 
The corresponding values for HeStar2 are 0.2~hrs (2.2~hrs) at $10^{5}$~yrs and 0.1~hrs (0.4~hrs) at $10^{7}$~yrs.

\section{Conclusions}
\label{sec:conclusions}

We have modeled the interaction between a He WD/ He star donor and SN Ia ejecta. 
We adopt an ejecta profile based on sub-$\mch$ explosion models \citep[][]{Shen2018_subMch}, and vary the explosion kinetic energy. 
During the interaction, $\approx 0.01 - 0.02\,\msun$ of He-rich material is stripped from the donor and mixed into the SN ejecta. The stripped donor mass loss peaks at $v \approx $~1,000\,$\kms$, and dominates over SN ejecta within a cone of half-opening angle $\approx 15^{\circ}$. 
The velocity distribution and the multi-dimensionality of the stripped material should be accounted for in future radiative transfer calculations \citep[e.g.,][]{Botyanszki2018}. 
Placing detection limits on the stripped He material from late-time SNe Ia spectrum could then constrain the donor type of the progenitor system \citep[e.g.,][]{Lundqvist2015}. 
We also find that $\sim 10^{-7} - 10^{-4}\,\msun$ of ejecta material, mostly consisting of iron-group elements, remains bound to the donor. However, we cannot accurately estimate the amount of bound ejecta material due to our finite box sizes.

Due to entropy deposition from the shock caused by the SN ejecta, the post-explosion donor appears bright and puffy for $\approx 10^{5}-10^{6}$~yrs. This is in contrast to \cite{Bhat2024}, who find that for CO WD donors, the post-explosion donor rapidly cools in $\lesssim 10^{3}$~yrs. We show that the post-explosion properties of our HeWD model, whose pre-explosion orbital velocity agrees well with the hypervelocity star D6-2 \citep[][]{Wong2023}, also agrees reasonably with D6-2. This strengthens our argument that D6-2 is a surviving low-mass ($\approx 0.1-0.2\,\msun$) He WD donor \citep[see also][]{Bauer2021}. However, we do not account for the surface metal pollution of D6-2, likely due to the capture of low-velocity SN ejecta. We defer this to future works.

Our {\athena} models are limited by the use of a $\Gamma=5/3$ EOS (However, see Appendices \ref{appendix_B} \& \ref{appendix_C}). We plan to use a realistic EOS accounting for radiation and degeneracy in future {\athena} simulations \citep[][]{Coleman2020}. 
This will open up the opportunity to investigate other donor types, including C/O WDs which likely comprise the majority of the seven hypervelocity stars \citep[][]{Shen2018_D6,Bauer2021,ElBadry2023}, and allow us to clarify the post-explosion evolution of the hypervelocity stars.

\section*{Acknowledgments}


We thank the referee for their constructive suggestions that have greatly improved our manuscript. 
We are grateful to Ken Shen for sharing his sub-$\mch$ explosion models and for conversations about the hypervelocity stars. 
We thank Evan Bauer for helpful conversations regarding $\mesa$ modelling and for sharing his He star models from \cite{Bauer2019}. 
We thank Logan Prust for insightful discussions regarding the {\athena} simulations. 
This research benefited from stimulating interactions with Kareem El-Badry and Jim Fuller. 
We thank Sterl Phinney for pointing out the need to account for orbital motion in the mass loss velocity distribution. 
This work made use of the Heidelberg Supernova Model Archive (HESMA), \href{https://hesma.h-its.org}{https://hesma.h-its.org}.
This work was supported, in part, by the National Science Foundation through grant PHY-2309135, and by 
the Gordon and Betty Moore Foundation through grant GBMF5076. 
The computations in this work were, in part, run at facilities supported by the Scientific Computing Core at the Flatiron Institute, a division of the Simons Foundation. 
Use was made of computational facilities purchased with funds from the National Science Foundation (CNS-1725797) and administered by the Center for Scientific Computing (CSC). The CSC is supported by the California NanoSystems Institute and the Materials Research Science and Engineering Center (MRSEC; NSF DMR 2308708) at UC Santa Barbara.


\software{
\texttt{Athena++} \citep[][]{Stone2020}, 
\texttt{MESA} \citep[v24.03.1;][]{MESAI,MESAII,MESAIII,MESAIV,MESAV,MESAVI}, 
\texttt{py\_mesa\_reader} \citep{bill_wolf_2017_826958},
\texttt{ipython/jupyter} \citep{perez_2007_aa,kluyver_2016_aa},
\texttt{matplotlib} \citep{hunter_2007_aa},
\texttt{NumPy} \citep{numpy2020}, 
\texttt{SciPy} \citep{scipy2020}, 
\texttt{Astropy} \citep{astropy:2013,astropy:2018},
and 
\texttt{Python} from \href{https://www.python.org}{python.org}
}

\newpage

\appendix

\twocolumngrid

\begin{deluxetable*}{ccccc|ccc}
\tablenum{3}
\tablecaption{Comparison of different numerical choices for HeStar2, $\Eke = 0.7 \times 10^{51}$~erg, $\Mej = 0.779\,\msun$}
\tablehead{
\colhead{Self-gravity solver} & 
\colhead{Resolution} & 
\colhead{Box size} &
\colhead{$\rho_{\rm floor}$} & 
\colhead{$P_{\rm floor}$} & 
\colhead{$M^{f}_{\rm He}$} & 
\colhead{$\rho^{f}_{\rm c}$} & 
\colhead{$v_{\rm kick,x}$}
\\
\colhead{} &
\colhead{} &
\colhead{[$x_{0}$]} &
\colhead{[$\rho^{i}_{\rm c}$]} &
\colhead{[$P^{i}_{\rm c}$]} &
\colhead{[$\msun$]} &
\colhead{[$\rho^{i}_{\rm c}$]} &
\colhead{[$\kms$]} 
}
\tablewidth{100pt}
\startdata
Fourier & $256^{3}$ & $(-10,50)$ & $10^{-5}$ & $1.65\times10^{-8}$ & 0.214 & 0.546 & 192 \\
Fourier & *$512^{3}$ & $(-10,50)$ & $10^{-5}$ & $1.65\times10^{-8}$ & 0.216 & 0.566 & 195  \\
Fourier & $512^{3}$ & *$(-15,65)$ & *$10^{-6}$ & *$ 10^{-8} $ & 0.220 & 0.561 & 178  \\
*multigrid & $512^{3}$ & $(-15,65)$ & *$10^{-5}$ & $ 10^{-8} $ & 0.228 & 0.594 & 232  \\
multigrid & $512^{3}$ & $(-15,65)$ & *$10^{-6}$ & $ 10^{-8} $ & 0.226 & 0.568 & 212  \\
multigrid & $512^{3}$ & $(-15,65)$ & *$10^{-7}$ & *$ 10^{-9} $ & 0.226 & 0.568 & 208  \\
\enddata
\label{tab:model2c_numerics}
\tablecomments{Asterisk (*) highlights changes relative to the previous row. The box size is for the $x$-axis. The $y$ and $z$-axes have the same length as the $x$-axis, but symmetric around the origin. Here $\rho_{\rm floor}$ and $P_{\rm floor}$ refer to the floor density and pressure, $M^{f}_{\rm He}$ and $\rho^{f}_{\rm c}$ refer to the final donor mass and central density, and $v_{\rm kick,x}$ is the kick velocity the donor receives in the $x$-direction. }
\end{deluxetable*}

\begin{deluxetable*}{cccc|ccc}
\tablenum{4}
\tablecaption{Comparison of different numerical choices for HeWD, $\Eke = 1.2 \times 10^{51}$~erg, $\Mej = 1.0\,\msun$}
\tablehead{
\colhead{Resolution} & 
\colhead{Box size} &
\colhead{$\rho_{\rm floor}$} & 
\colhead{$P_{\rm floor}$} & 
\colhead{$M^{f}_{\rm He}$} & 
\colhead{$\rho^{f}_{\rm c}$} & 
\colhead{$v_{\rm kick,x}$}
\\
\colhead{} &
\colhead{[$x_{0}$]} &
\colhead{[$\rho^{i}_{\rm c}$]} &
\colhead{[$P^{i}_{\rm c}$]} &
\colhead{[$\msun$]} &
\colhead{[$\rho^{i}_{\rm c}$]} &
\colhead{[$\kms$]} 
}
\tablewidth{100pt}
\startdata
$512^{3}$ & $(-15,85)$ & $10^{-6}$ & $10^{-8}$ & 0.108 & 0.211 & 218 \\
$704^{3}$ & $(-15,85)$ & $10^{-6}$ & $10^{-8}$ & 0.106 & 0.196 & 230  \\
$896^{3}$ & $(-15,85)$ & $10^{-6}$ & $10^{-8}$ & 0.106 & 0.191 & 240  \\
$704^{3}$ & $(-15,85)$ & $3 \times 10^{-7}$ & $ 3 \times 10^{-10}$ & 0.106 & 0.194 & 219  \\
\enddata
\label{tab:HeWD_numerics}
\tablecomments{ Variables take the same meaning as Table \ref{tab:model2c_numerics}. }
\end{deluxetable*}

\section{Numerical choices and comparison to Bauer et al. 2019}
\label{appendix_A}

We attempted to reproduce the results of \cite{Bauer2019} for their donor model 2 (HeStar2 here) with $\Eke = 0.7 \times 10^{51}$~erg, $\Mej = 0.779\,\msun$. 
We retain the same simulation box size, spatial resolution ($256^{3}$ cells), density and pressure floors, donor relaxation process, ejecta profile \citep[power-law;][]{Kasen2010}, 
and self-gravity solver (Fourier method). However, we do not recover the same results. For example, their final donor has $M^{f}_{\rm He} \approx 0.15\,\msun$ and $ \rho^{f}_{\rm c} \approx 0.25 \rho^{i}_{\rm c}$ (see their Table 2), while we find $M^{f}_{\rm He} \approx 0.21\,\msun$ and $ \rho^{f}_{\rm c} \approx 0.55 \rho^{i}_{\rm c}$. This means that we find much less mass loss, given that the pre-explosion mass of HeStar2 is $0.233\,\msun$, and we find a more tightly bound post-explosion donor. While the difference in $M^{f}_{\rm He}$ could be due to different definitions for bound material, $\rho^{f}_{\rm c}$ should only depend on the shock strength and hence numerics such as spatial resolution. We were not able to resolve these differences. 

We explored the effects of a number of numerical choices for the above simulation. 
The results are detailed in Table \ref{tab:model2c_numerics}. 
We find that the resolution in \cite{Bauer2019} is likely sufficient, but the box size is likely too small, leading to the loss of some bound mass. 
We also find that a high density floor can induce a net gravitational pull given a simulation domain that is asymmetric around the $x$-axis. 
However, all of these choices do not change $\rho^{f}_{\rm c}$ by more than 10\%, and the donor mass loss only ranges from $0.01-0.02\,\msun$, and so cannot explain the discrepancy with \cite{Bauer2019}. 
For our main results in the paper, we have varied resolution, box size, and floor limits to ensure that our results are reasonably converged.  This leads us to adopt the numerical choices detailed in Section \ref{sec:numerics}. An example for the HeWD, $\Eke = 1.2 \times 10^{51}$~erg model is shown in Table \ref{tab:HeWD_numerics}. We do not show the bound ejecta mass because all choices lead to a change by factor of a few around $10^{-5}\,\msun$ and this quantity is limited by our finite box size anyways.

\section{Mapping between {\mesa} and {\athena}}
\label{appendix_B}

\begin{figure}
    \centering
    \includegraphics[width=\linewidth]{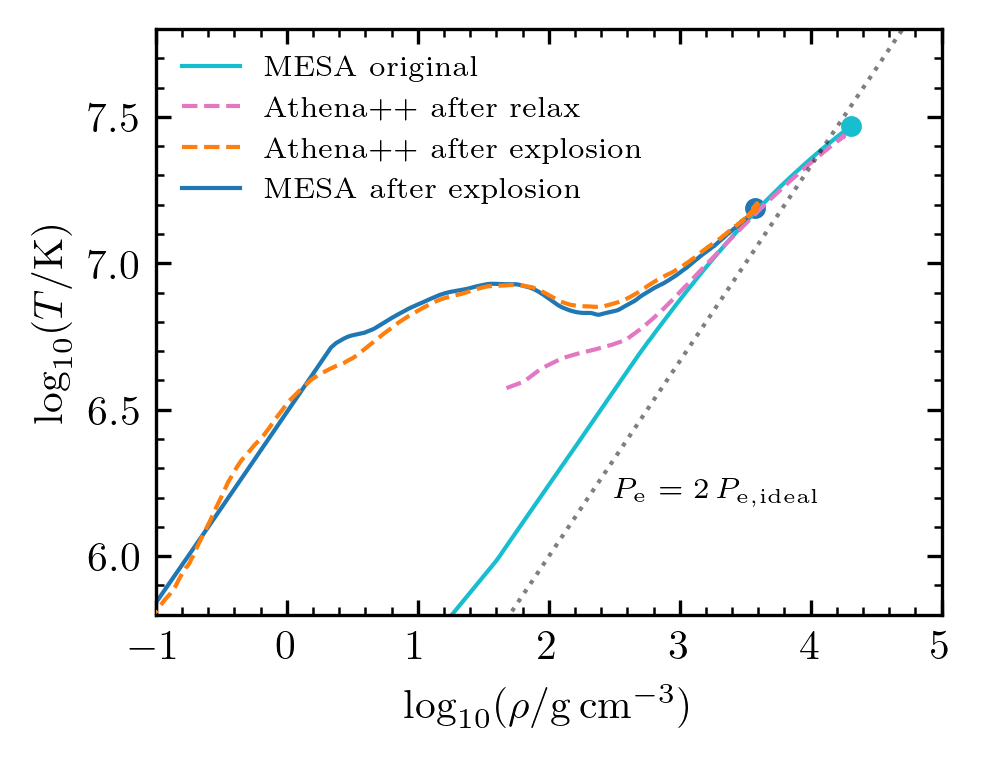}
    \caption{ Temperature-density profiles of the HeWD donor. 
    We show the original {\mesa} profile (light-blue solid), the {\athena} profile after relaxation (pink dashed), the {\athena} profile near the end of the simulation (orange dashed), and the {\mesa} that is then further evolved (blue solid). The dotted grey line shows the transition to degeneracy, where the electron pressure is twice the electron ideal gas pressure. The circle markers show the center of the models. }
    \label{fig:TRho_relax}
\end{figure}

To quantify the errors introduced by adopting a $\Gamma=5/3$ EOS, we compare the temperature and specific entropy profiles of the donor star between {\mesa} and {\athena}, for the HeWD, $\Eke=1.2\times 10^{51}$~erg run. To obtain the {\athena} profiles, we take spherical averages of $P$ and $\rho$, and solve for $T$ using the {\mesa} \texttt{eos} module. 
The $T-\rho$ profiles are shown in Figure \ref{fig:TRho_relax}. 
We find good agreement between the original {\mesa} profile and the {\athena} profile after the relaxation as detailed in Section \ref{sec:roche} (but before explosion), except near the outermost layers which is affected by interactions with floor material, though this material is most likely stripped during the explosion. 
There is also good agreement between the {\athena} profile at the end of the simulation (after explosion), and the {\mesa} profile that is then evolved further (see Section \ref{sec:postexp}). 
We note that the original {\mesa} profile is only semi-degenerate in its center, which is why we adopted $\Gamma=5/3$. We also note 
the drop in center density after the explosion which indicates expansion of the donor. 

\begin{figure}
    \centering
    \includegraphics[width=\linewidth]{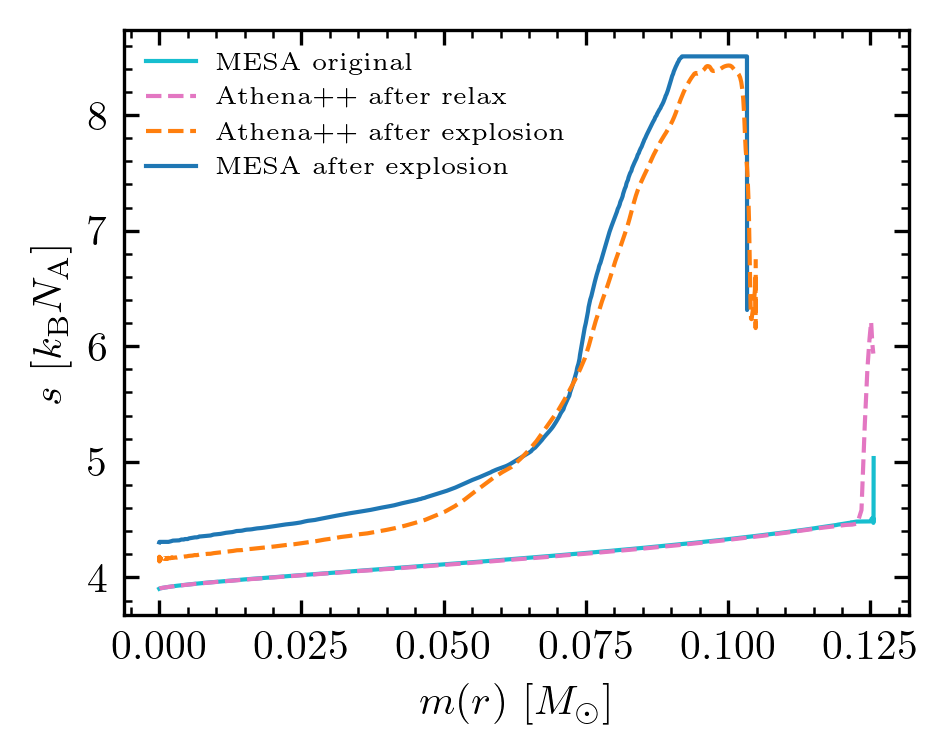}
    \caption{ Entropy profiles of the HeWD donor. The lines are the same as Figure \ref{fig:TRho_relax}. }
    \label{fig:entropy}
\end{figure}

We also use the {\athena} $P-\rho$ profiles to obtain the specific entropy profile, by adding the original specific entropy profile from {\mesa} and the change in specific entropy $\Delta s$ from {\athena} (see eqn. \ref{eqn:entropy}). The resulting profiles are shown in Figure \ref{fig:entropy}. 
Although there is a slight offset in the center between {\mesa} and {\athena} profiles which reflects the error in using a $\Gamma=5/3$ EOS, the overall agreement is still good.

\section{Realistic equation of state}
\label{appendix_C}

\begin{figure}
    \centering
    \includegraphics[width=\linewidth]{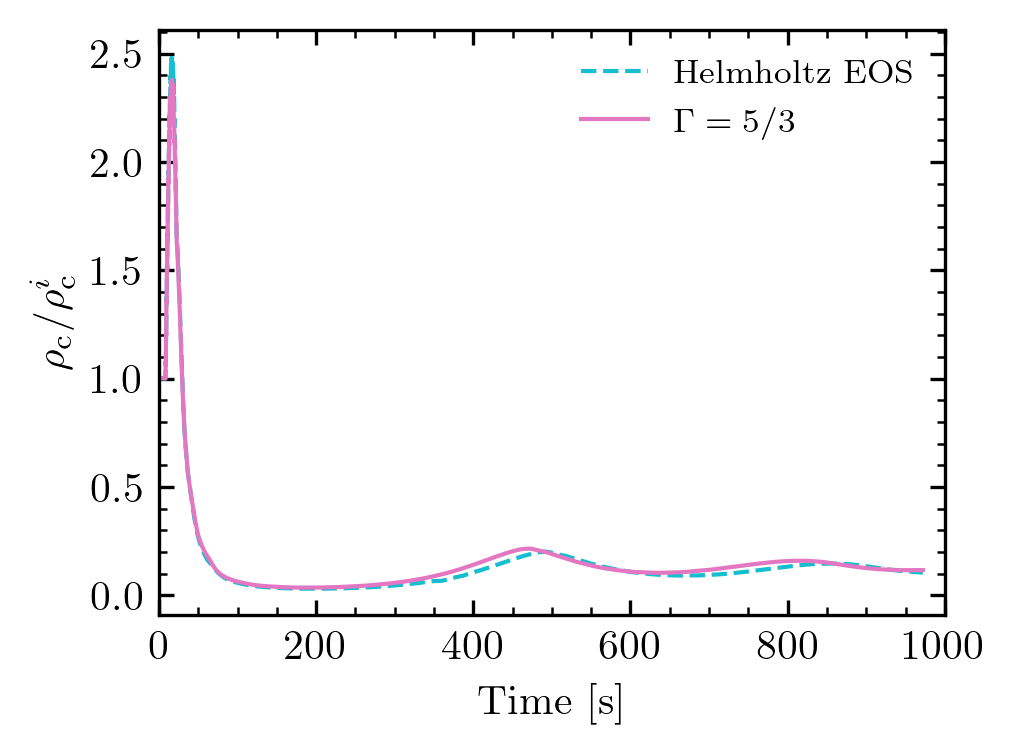}
    \caption{ Central density of the HeWD donor for $\Gamma=5/3$ (solid pink) and the Helmholtz EOS (blue dashed). }
    \label{fig:helm_rho}
\end{figure}

After our first submission, we began experimenting with using a realistic EOS using {\athena}'s general EOS capability \citep[][]{Coleman2020}. 
We compare two runs with the HeWD, $\Eke = 1.2 \times 10^{51}$~erg model, one with $\Gamma = 5/3$ and one with the Helmholtz EOS \citep[][]{Timmes2000}, which accounts for radiation, electron degeneracy and Coulomb corrections. For this initial experimentation, we assume a pure He composition for all components including the ejecta, and do not relax the donor star under a Roche potential. 
The donor central density during the simulation is shown in Figure \ref{fig:helm_rho}, which shows a similar evolution between the two EOS's. Other quantities are also similar, such as the final bound mass $M^{f}_{\rm He}$ which only differs by $\approx 2\%$ between the two. We will present more details on our use of a realistic EOS in a future paper, but this initial comparison justifies our use of $\Gamma=5/3$.

\bibliography{athena,main,mesa,software}{}
\bibliographystyle{aasjournal}

\end{document}